# Experimentally correlating thermal hysteresis and phase compatibility in multifunctional Heusler alloys

A.A. Mendonça[1], L. Ghivelder[1], P.L. Bernardo[2], Hanlin Gu[3], R.D. James[3], L.F. Cohen[4], and A.M. Gomes[1]

[1] Instituto de Física, Universidade Federal do Rio de Janeiro, Cidade Universitaria, Rio de Janeiro 21941-972, Brazil

[2] Centro Brasileiro de Pesquisas Físicas, Rua Dr. Xavier Sigaud, 150, Rio de Janeiro, Rio de Janeiro, 22290-180, Brazil

[3] Aerospace Engineering and Mechanics, University of Minnesota, USA

[4] Blackett Laboratory, Imperial College London, South Kensington Campus, London, SW7 2BZ, United Kingdom

## Abstract

Thermal hysteresis is recognized as one of the main drawbacks for cyclical applications of magnetocaloric and ferromagnetic shape memory materials with first order transformations. As such, the challenge is to develop strategies that improve the compatibility between the phases involved in the transitions and study its influence on thermal hysteresis. With this purpose, we explore the thermal, structural and magnetic properties of the $Ni_2Mn_{1-x}Cu_xGa_{0.84}Al_{0.16}$ Heusler alloys. The alloys present a thermal hysteresis reduction of ~ 60% when the Cu content in the compound varies from x = 0.10 to x = 0.25, with a minimum hysteresis width of 6 K being achieved. We applied the geometric non-linear theory of martensite to address the phase compatibility, quantified by the parameter $\lambda_2$, the middle eigenvalue of the transformation stretch tensor, and found that the minimum of hysteresis is associated with a better crystallographic compatibility ($\lambda_2$ closer to 1) between the austenite and martensite phases. In addition, we show that the valley-like properties of hysteresis found in the $Ni_2Mn_{1-x}Cu_xGa_{0.84}Al_{0.16}$ compounds is present in several other alloys in the literature. These results provide new pathways to understand as well as to masters the phase compatibility and ultimately achieve a low thermal hysteresis in multifunctional Heusler alloys.

**I. Introduction**

Compounds with a magnetic-field-induced first order transition are widely studied for future use in technological devices. Among them, $Gd_5(Si,Ge)_4$,[1] $MnFeP_{1-x}As_x$,[2] $La(Fe,Si)_{13}$[3] and $Ni_2MnQ$-based (Q = Ga, In, Sn, Sb) Heusler alloys[4,5] are promising candidates for applications in magnetic refrigeration. The latter are also used in smart actuators, energy harvesting and robotics applications. The main phenomena investigated in these materials are the magnetocaloric effect (MCE) and the ferromagnetic shape memory effect. The MCE is a temperature change observed when the magnetic field applied to sample is varied. The ferromagnetic shape memory effect is a magnetic-field-induced deformation that causes strain or stress to the sample. Both effects play a major role for application purposes. However, there are some outstanding difficulties, in particular, how to overcome magnetic and thermal hysteresis. A large hysteresis prevents the use of low magnetic fields for practical use.[6,7] In addition, a correlated problem is the sample's structural instability when submitted to magnetic/thermal cycles, which will induce cracks and fatigue inside the material if there is a large volume change at the structural transition.[7,8]

Although $Ni_2MnGa$-based alloys are promising for applications due to giant MCE and large ferromagnetic shape memory effect, thermal hysteresis is considerable, of the order of tens of Kelvin, which prevents cyclical applications.[9,10] Usually, these Heusler alloys crystallize in a cubic phase and present a martensitic transition when cooled. The formation of a martensitic microstructure and its geometric compatibility with the higher temperature austenite phase is reported as a way to understand the hysteresis phenomenon.[8] Composition changes as well as external stimuli have been used previously to decrease the hysteresis.[7,11,12]

Both magnetocaloric and ferromagnetic shape memory applications require the application of a magnetic field. In addition, a significant magnetization difference between the structural phases is required for the structural transition to be triggered by the applied field.[13,14] Some particular compositions of Heusler alloys present a simultaneous change of structural symmetry and magnetic ordering, known as a magnetostructural transition (MST).[12,15] In this case, a high MCE occurs between the paramagnetic austenite (parent) and ferromagnetic martensite (lower temperature and symmetry) phases.[15,16] In contrast, Heusler alloys with ferromagnetic austenite phase and antiferromagnetic martensite phase, give rise to what is known as an inverse MCE.[9,12] For ferromagnetic shape memory effect applications there are two main mechanisms to generate the required strain/stress: the martensitic direct-reverse transformation, also called metamagnetic shape memory effect,[12,13,17] and the reorientation of the martensite variants by twin boundary motion.[4,18,19]

The geometric non-linear theory of martensite addresses the microstructure of martensite as well as its formation.[20,21] To this end, it makes use of crystalline symmetries and geometric compatibility of the phases involved. Since these subjects are extremely well correlated to the hysteresis within the transformation, the theory can help us understand the mechanisms to decrease this undesired property of martensitic transitions. Basically, the aim is to quantify the geometric compatibility between the phases and, consequently, achieve a thermal hysteresis as low as possible. In addition, since hysteresis is related to dissipated work, the optimized phase compatibility is also expected to prevent the creation of defects. Hence, an improvement of the phase compatibility leads to higher resistance to fractures.

In a few words, the geometric non-linear theory of martensite takes two conditions to search for the minimum of the hysteresis: (i) $\det(U) = \lambda_1\lambda_2\lambda_3 = 1$, where U is the transformation stretch tensor of the transformation; (ii) $\lambda_2 = 1$, where $\lambda_1$, $\lambda_2$ and $\lambda_3$ are the eigenvalues of U, in crescent order. The tensor U is determined by means of the lattice parameters of the unit cells from martensite and austenite phases. If $\det(U) = 1$, the volume change between the phases under transformation is zero. Moreover, the $\lambda_2$ parameter measures the compatibility of the austenite with a single variant of the martensite, where the condition $\lambda_2 = 1$ means that they are fully compatible. Therefore, by mapping $\lambda_2$ under composition changes in an alloy one can determine the concentration that maximizes the phase compatibility and reach a lower thermal hysteresis. Earlier studies have applied the geometric nonlinear theory of martensite to identify compounds with lower hysteretic behaviour.[20,21] Quite recently this approach was also exploited in studies of Ni-Mn-In Heusler alloys.[22,23]

The Ni$_2$MnGa compound crystallizes in a cubic L2$_1$-type structure, space group Fm-3m, with lattice parameter $a$ = 5.825 Å at room temperature and presents a lower temperature martensitic structure.[24] It shows a second order paramagnetic to ferromagnetic transition at 376 K and a martensitic transition around 200 K.[9] Small substitution of Ni on the Mn site leads to a MST with a giant MCE around 333 K.[25] Previously, it was shown that it is possible to couple the structural and magnetic transition to achieve a giant MCE in the Ni$_2$Mn$_{1-x}$Cu$_x$Ga alloy as well.[26] The replacement of Mn by Cu yields a MST at 308 K for x = 0.25. In these alloys, an important parameter associated with achieving a MST is the number of valence electrons per atom (*e/a*). In general, if an element with larger *e/a* than some precursor of the Ni$_2$MnGa-based alloy is substituted (without generating considerable change in the lattice parameters or hybridization), the temperature of the martensitic transition increases.[27] Furthermore, Mn replacement by non-ferromagnetic elements decreases the magnetic transition temperature because of the relevant role of the indirect exchange interactions presented by the 3d electrons of Mn. This is the main mechanism responsible for the ferromagnetism within the alloy.[28] In this case, Cu (4s$^1$3d$^{10}$) substitution on the Mn (4s$^2$3d$^5$) site satisfies these requirements due to the larger *e/a* ratio of the Cu and non-ferromagnetic property of this 3d metal.

Further considerations are related to cost, as it is desirable to replace the Ga element by a cheaper one, and for this additional reason Al is attractive.[29] Aluminium substitution in Ni$_2$MnGa$_{1-x}$Al$_x$ alloys generates a coexistence of L2$_1$-type (ferromagnetic) and B2-type (antiferromagnetic) cubic structures in the parent phase, leading predominantly to antiferromagnetism when x > 0.30.[30,31] The fabrication and annealing processes also have influence in the alloy's magnetic and structural properties, as well as the ability to achieve a predominantly ferromagnetic L2$_1$-type structure.[30] The alloys Ni$_2$Mn$_{1-x}$Cu$_x$Ga$_{0.9}$Al$_{0.1}$ were previously studied.[32] In this family of compounds, the MST occurs around 295 K for x = 0.20, resulting in a magnetic entropy change of ΔS$_M$ = - 9.5

J·Kg$^{-1}$·K$^{-1}$ under 0-5 T with a 26% reduction in cost compared to Ni$_2$Mn$_{1-x}$Cu$_x$Ga and with an equivalent refrigerant capacity.

In this paper, by means of thermal, structural and magnetic measurements in Ni$_2$Mn$_{1-x}$Cu$_x$Ga$_{0.84}$Al$_{0.16}$ alloys, we identify a valley-like behavior in the Cu content dependence of the thermal hysteresis. We demonstrate that this behavior is present in other Heusler alloys that undergo a magnetostructural transition triggered by some composition change. In addition, we apply the geometric non-linear theory of martensite in our material to further investigate the phase transformation and better understand the evolution of the phase compatibility as the composition changes. The results show that, for these Heusler alloys and probably for other compounds, a specific compositional change leads to a minimum of thermal hysteresis. By constructing a phase diagram, it is possible to identify the composition with better relation of signal output for a given energy loss.

## II. Experimental details

Heusler alloys Ni$_2$Mn$_{1-x}$Cu$_x$Ga$_{0.84}$Al$_{0.16}$, with $x$ = 0.10, 0.20, 0.25, 0.30, 0.31, 0.35 and 0.45 were fabricated using conventional arc melting in 99.999% pure argon atmosphere and metallic elements of purity greater than 99.99%. During the initial fabrication process, we noticed the Mn loss to be approximately 3%, and to account for this, subsequent processes included a 3% Mn excess before the melting to ensure the correct final stoichiometry. To achieve greater homogenization, two thermal treatments were applied, in which the samples were wrapped with tantalum foil and encapsulated in quartz tubes under a low argon pressure of 0.2 atm. The first thermal annealing was for 72 h at 1273 K and the second for 24 h at 673 K using a temperature ramp of 3 K/min. The sample was quenched at room temperature with water at the end of each annealing process. In order to verify the final composition, electron dispersive spectroscopy (EDS) measurements were made in a JEOL 7100FT scanning electron microscope and an Oxford Nordlys 80 detector. The EDS results are presented in Table I. The data clearly show the Mn replacement by Cu in the series. The measured Mn concentration is larger than the nominal values probably because an excess of Mn added initially to overcome losses in the arc melting process. X-ray powder diffraction (XPD) data of all samples were collected on a X'Pert Pro (PANalytical) X-ray diffractometer using the Bragg-Brentano geometry with 2θ of 0.20° and CuKα1 radiation of λ = 1.54056 Å. The crystal structure was characterized by Le Bail analysis of the XPD data, using the FullProf software suite. The reliability factors of the analysis are within the range $5.53 \leq R_p \leq 9.97$ and $4.78 \leq R_{wp} \leq 7.94$. Heat flow measurements using a Differential Scanning Calorimeter (DSC) Q2000 from TA Instruments Inc. were performed following a heat/cool/heat procedure at 10 K/min. Magnetization measurements were made as a function of temperature and magnetic field with a Vibrating Sample Magnetometer (VSM) in the Physical Properties Measurement System (PPMS) from Quantum Design Inc. The isofield magnetization measurements followed zero field cooled (ZFC) and field cooled cooling (FCC) processes at 1 K/min. Isothermal magnetization measurements were performed up to 9 T at 5 mT/sec constant rate. Dilatometry measurements as a function of temperature and applied magnetic field were made in the silver based Capacitance Dilatometer,[33] following a 0.2 K/min and 3 mT/s temperature and magnetic field sweep rate protocol, respectively.

## III. Results and Discussion

*A. Heat flow within the transitions*

Heat flow measurements in a differential scanning calorimeter (DSC) are ideal to study thermal properties of first order transitions. Similar to the $Ni_2Mn_{1-x}Cu_xGa$ alloys,[26] in the $Ni_2Mn_{1-x}Cu_xGa_{0.84}Al_{0.16}$ compounds we observe an increase of the structural transition temperature when the Cu concentration is increased, as shown in Fig. 1. The sample with x = 0.25 exhibits a thermo-elastic intermartensitic transition, sometimes present in other Heusler alloys.[34,35] When the austenitic phase transforms into the martensitic one and then another structural transformations occurs between martensitic phases, the structural change is called intermartensitic transition. In general, these transformations in $Ni_2MnGa$-based materials occurs among modulated martensitic structures (5M and 7M), or among modulated and a non-modulated tetragonal structure.[15,34]

As seen clearly in the martensitic transition for x = 0.10, 0.20, 0.25, 0.31 and in the austenitic transition for x = 0.20 and 0.30, some transformations present multiple peaks.

The enthalpy of the transition as a function of Cu content, calculated from the heat flow measurements for the first order transitions, is presented in the inset of Fig. 1. There is a boost to the total enthalpy where the value of enthalpy exceeds the curve of other samples. This behavior is due to the magnetostructural coupling in the samples $x = 0.30$ and $x = 0.31$.

*B. Magnetization and Dilatometry*

Fig. 3 shows the zero-field cooled (ZFC) magnetization measurements for the alloys with $x = 0.20, 0.25, 0.30, 0.31, 0.35$ and $0.45$, measured with an applied magnetic field of 20 mT. The samples with Cu concentration $x = 0.10, 0.20$ and $0.25$ show second order magnetic transitions between paramagnetic and a ferromagnetic phase whereas the lower temperature transitions are structural transformations. As the Cu content increases, the structural transition moves up and the magnetic transition moves down in temperature. Then, for $x = 0.30$ and $0.31$, the magnetostructural transformation appears. Now, the transition from a paramagnetic to a ferromagnetic phase is controlled by the structural change, thereby transforming as a first order transition. Therefore, these samples present an abrupt magnetic ordering change, which coupled with the structural transition is the cause of the jump in the enthalpy exhibited inset the Fig. 1. On the other hand, the compositions $x = 0.35$ and $0.45$ show only a second order magnetic transition and no structural transformation is observed since it occurs between paramagnetic phases.

*C. Phase diagram*

The phase diagram shown in Fig. 3 was constructed using heat flow and magnetization results. From the DSC data, the mean martensitic and austenitic temperatures were calculated as $(T_{MS}+T_{MF})/2$ and $(T_{AS}+T_{AF})/2$, respectively, where the index S represents the start and F the final temperature of the transitions. In addition, the Curie point was obtained as the inflection point of the magnetization data. The values obtained are shown in Tab. II. As noticed before, the substitution of Mn by Cu increases the martensitic transition temperature, while lowering the magnetic transition temperature. These substitution effects are mainly due to the larger e/a relation of the Cu and the partial Mn substitution by a non-ferromagnetic element, respectively, as already discussed.

For samples with x = 0.10, 0.20 and 0.25, the magnetic transition occurs in the austenite phase. For these compositions, the Cu content dependence of both structural and magnetic transition temperatures presents a linear behavior. With x = 0.30 and x = 0.31, a discontinuous magnetic ordering change occurs at the structural transition, due to the MST, and we notice an increase of the slope for the structural transformation temperature. For x ≥ 0.35, this slope rises further. In addition, in this interval the negative slope of the magnetic transition is even smaller, now in the martensite phase. By observing the phase diagram, we note that the alloy presents a MST from x = 0.28 to x = 0.33.

*D. Composition dependence of the hysteresis*

In Fig. 4, we show the thermal hysteresis width, $\Delta H$, as a function of the Cu content in the alloys. The results were obtained from heat flow, magnetization and thermal expansion measurements (not shown) and were calculated using the transition temperatures of the phase diagram. The data from these different techniques are in close agreement. The results show a minimum in $\Delta H$ as a function of the Cu content. This clearly distinguishes three different regions: (i) $\Delta H$ decreases when Cu is added in the alloys up to x = 0.25, which corresponds to region 1 of the phase diagram, where both martensite and austenite phases are ferromagnetic at temperatures close to the structural transition. (ii) For x = 0.30 and x = 0.31, which corresponds to the region 2, the thermal hysteresis increases. Here, the austenite phase becomes paramagnetic while the martensite remains ferromagnetic. (iii) For higher Cu compositions, the region 3 where both phases are paramagnetic at temperatures close to the structural transformation, $\Delta H$ still increases but it tends to stabilize. By comparing the phase diagram of Fig. 3 and the hysteresis plot of Fig. 4, we conclude that the composition with a MST and smaller thermal hysteresis is close to x = 0.28, with thermal hysteresis around 7.2 K. The compound with this Cu content is expected to be the best one among our series of samples in the relation power output per energy loss and cyclical fatigue at the transformation.

Since the tuning of the thermal hysteresis plays an important role in the development of materials attractive for technology, the observed minimum in $\Delta H$ along with a detailed study of the alloy properties can help facilitate further understanding of how to improve different materials for applications. In order to address this potential for applications, we measured cycles of heat flow around the temperature of the structural transitions for x = 0.25 and 0.30, as seen in the Fig. 5. In terms of reproducibility, the result shows a better performance for the martensitic transformation of the sample x = 0.25, while its intermartensitic transition and the magnetostructural transformation for x = 0.30 present larger instability. This signals that the smaller the hysteresis the better it is the reproducibility of the transformation and, then, it points to a correlation of transition reproducibility and thermal hysteresis width. The geometric non-linear theory of martensite explains that by lowering thermal hysteresis of the transformation, leads to improved cyclical efficiency, therefore less cracks are observed in under cyclical procedures.[20] The higher reproducibility of the material x = 0.25 seems to corroborate this statement.

*E. X-ray diffraction*

We concentrate our X-ray powder diffraction measurements on x = 0.10, 0.20, 0.25, 0.30 and 0.31 because the other compositions present a transition temperature above the instrumental range available in the XRD measuring system. We are interested in identifying the phases as well as obtain the lattice parameters of each phase involved in the first order transformation, in order to apply the geometric non-linear theory of martensite. Fig. 6 presents typical XPD results on the x = 0.10 sample. Results at 150 K correspond to a single martensite phase, at 310 K to an austenite phase, and closer to the transition temperature, at 204 K, both phases coexist. The figure shows the measured diffractograms, the calculated refined curves, Bragg positions for each phase, and the difference between the observed and calculated data.

Results around room temperature for all the measured samples (not shown) present a cubic L2$_1$, space group Fm-3m, the same austenitic structure of the Ni$_2$MnGa material.[24] At lower temperatures, the samples with x = 0.10, 0.20, and 0.25 present a monoclinic martensitic structure, space group I12/ma, also called 5M (five-layered). In addition, the sample with x = 0.25 also presents an intermartensitc transition, as observed in the DSC measurements of Fig. 1. This compound transforms from cubic L2$_1$ to the 5M martensite phase in the martensitic transition and transforms from 5M to the non-modulated L1$_0$ tetragonal structure at the intermartensitic transformation. On the other hand, for x = 0.30 and 0.31, the material transforms directly from cubic L2$_1$ to the non-modulated L1$_0$ tetragonal phase at the martensitic transition. These changes of the martensite phase as a function of the Cu composition were reported as being related to the number of valence electrons per atom, e/a. As the e/a increases, the alloy tends to move away from the cubic-to-modulated martensitic transformation and evolve into the cubic-to-non-modulated transformation.[36] The non-modulated structure is considered a ground state, since for this structure an intermartensitic transition is not observed.[34]

*E. Phase compatibility*

In order to obtain the phase compatibility in the samples, we applied the geometric non-linear theory of martensite,[20,21] using the XRD refined data. Crystallographic compatibility is quantified by the parameter $\lambda_2$, the middle eigenvalue of the transformation stretch. The Monoclinic II lattice has a unique 2-fold axis along an edge of the original cubic lattice, so the variants are also called "cube-edge" variants.[37] The number of variants for this structure is 12, therefore are 12 transformation stretch tensors. Given the difficulty to find detailed information about these stretch tensors and their respective eigenvalues in the literature, we believe it might be useful to present them here:

$$U_1 = \begin{pmatrix} \beta & 0 & 0 \\ 0 & \rho & \sigma \\ 0 & \sigma & \tau \end{pmatrix}, U_2 = \begin{pmatrix} \beta & 0 & 0 \\ 0 & \rho & -\sigma \\ 0 & -\sigma & \tau \end{pmatrix},$$

$$U_3 = \begin{pmatrix} \beta & 0 & 0 \\ 0 & \tau & \sigma \\ 0 & \sigma & \rho \end{pmatrix}, U_4 = \begin{pmatrix} \beta & 0 & 0 \\ 0 & \tau & -\sigma \\ 0 & -\sigma & \rho \end{pmatrix},$$

$$U_5 = \begin{pmatrix} \rho & 0 & \sigma \\ 0 & \beta & 0 \\ \sigma & 0 & \tau \end{pmatrix}, U_6 = \begin{pmatrix} \rho & 0 & -\sigma \\ 0 & \beta & 0 \\ -\sigma & 0 & \tau \end{pmatrix},$$

$$U_7 = \begin{pmatrix} \tau & 0 & \sigma \\ 0 & \beta & 0 \\ \sigma & 0 & \rho \end{pmatrix}, U_8 = \begin{pmatrix} \tau & 0 & -\sigma \\ 0 & \beta & 0 \\ -\sigma & 0 & \rho \end{pmatrix},$$

$$U_9 = \begin{pmatrix} \rho & \sigma & 0 \\ \sigma & \tau & 0 \\ 0 & 0 & \beta \end{pmatrix}, U_{10} = \begin{pmatrix} \rho & -\sigma & 0 \\ -\sigma & \tau & 0 \\ 0 & 0 & \beta \end{pmatrix},$$

$$U_{11} = \begin{pmatrix} \tau & \sigma & 0 \\ \sigma & \rho & 0 \\ 0 & 0 & \beta \end{pmatrix}, U_{12} = \begin{pmatrix} \tau & -\sigma & 0 \\ -\sigma & \rho & 0 \\ 0 & 0 & \beta \end{pmatrix},$$

where

$$\rho = \frac{\alpha^2 + \gamma^2 + 2.\alpha.\gamma(\sin\theta + \cos\theta)}{2\sqrt{\alpha^2 + \gamma^2 + 2.\alpha.\gamma.\sin\theta}},$$

$$\sigma = \frac{\alpha^2 - \gamma^2}{2\sqrt{\alpha^2 + \gamma^2 + 2.\alpha.\gamma.\sin\theta}},$$

$$\tau = \frac{\alpha^2 + \gamma^2 + 2.\alpha.\gamma(\sin\theta - \cos\theta)}{2\sqrt{\alpha^2 + \gamma^2 + 2.\alpha.\gamma.\sin\theta}},$$

and

$$\beta = \frac{b}{a_0}, \alpha = \frac{\sqrt{2}a}{a_0}, \gamma = \frac{\sqrt{2}c}{n.a_0}, \text{ with } n = 5.$$

In a cubic to tetragonal transformation there are 3 variants and the transformation stretch tensors are:

$$U_1 = \begin{pmatrix} \beta & 0 & 0 \\ 0 & \alpha & 0 \\ 0 & 0 & \alpha \end{pmatrix}, U_2 = \begin{pmatrix} \alpha & 0 & 0 \\ 0 & \beta & 0 \\ 0 & 0 & \alpha \end{pmatrix}, U_3 = \begin{pmatrix} \alpha & 0 & 0 \\ 0 & \alpha & 0 \\ 0 & 0 & \beta \end{pmatrix},$$

where

$$\alpha = \frac{\sqrt{2}a}{a_0} \text{ and } \beta = \frac{c}{a_0}$$

With the expression of the transformation stretch tensor $U_i$ of the cubic to monoclinic and cubic to tetragonal transformations, and the cell parameters obtained from the XRD data, we obtain the middle eigenvalues, $\lambda_2$, as displayed in Table II. In Fig. 7 we plot the values of the thermal hysteresis (see Fig. 4) as a function of the obtained $\lambda_2$ for each composition studied. These results clearly show that a lower thermal hysteresis is obtained for materials with values of $\lambda_2$ closer to one, which is consistent to the theory and previous investigations.[8,20,38] In addition, we demonstrate that Cu substitution on the Mn site in Ni(Mn,Cu)GaAl improves the crystallographic compatibility, while keeping a monoclinic transformation, which yields a minimum of hysteresis of the transition at x ~ 0.25. With higher Cu content, where the alloy presents a cubic to tetragonal transition, both thermal hysteresis increases and $\lambda_2$ grows beyond 1. Therefore, the thermal hysteresis behavior shown in Fig. 4 can be unambiguously attributed to the manipulation of the structural phase compatibility, promoted by a compositional substitution within the alloys.

*F. Valley of thermal hysteresis in the literature*

We used the previously defined transition temperatures $T_{MS}$, $T_{MF}$, $T_{AS}$ and $T_{AF}$ of some Heusler alloys reported in the literature to obtain their thermal hysteresis as a function of the Mn substitution. We analyzed previously published data on seven well-know alloys: the full Heusler alloys $Ni_2Mn_{1-x}Cu_xGa$ (Fig. 8a, Ref. 39); $Ni_{2+x}Mn_{1-x}Ga$ (Fig. 8b, Refs. 25, 40) and $Ni_2Mn_{1-x}Cr_xGa$ (Fig. 8c, Ref. 27) as well as the half Heusler alloys with different substitutions: $NiMn_{1-x}In_x$ (Fig. 8d, Ref. 5), $NiMn_{1-x}Al_x$ (Fig. 8e, Ref. 41) and $NiMn_{1-x}Sn_x$ (Fig. 8f, Ref. 42), which are $NiMn_{1-x}Q_x$-based alloys with Q = In, Al or Sn, respectively; and the full Heusler $Ni_{50.5}Mn_{25-x}Fe_xGa_{24.5}$ (Table I, Ref. 43), a quaternary $Ni_2MnGa$-based alloy with Mn replacement by ferromagnetic Fe.

As observed in the different plots of Fig. 8, except for the $Ni_2Mn_{1-x}Fe_xGa$, all other Heusler alloys show a minimum point in the composition dependence of the thermal hysteresis, which confirms that the scope of our results, related to the minimum of the thermal hysteresis, is broader than the specific alloy studied here. In the case of $Ni_2Mn_{1-x}Fe_xGa$, the thermal hysteresis increases linearly with the Fe content. However, this alloy is the only one listed in which the substitution does not converge to a magnetostructural transition at some temperature, i.e. this substitution only further separates the martensitic and magnetic transition temperatures. Mn replacement by Fe decreases the martensitic transition temperature (around 200 K for x = 0) and increases the magnetic one (around 376 K for x = 0). In all the other listed alloys, one specific characteristic repeats: while replacing Mn for some other element, the thermal hysteresis decreases before the concentration with MST and there is an increase of the thermal hysteresis when the substitution reaches the compositions with MST. This valley-like behavior provides a mechanism to estimate the best composition to achieve maximum efficiency in each alloy, by knowledge of its structural/magnetic phase diagram.

The minimum in the thermal hysteresis width for different materials occurs in a certain substitution interval whenever both structural phases are ferromagnetic or paramagnetic. Therefore, we cannot determine which type of magnetic ordering is best for structural phase compatibility. Nevertheless, on average, the minimum of $\Delta H$ is smaller for the Full Heusler alloys, which present transitions between ferromagnetic phases. Conversely, it is clear that the thermal hysteresis increases when the phases presents different type of magnetic ordering. Since the hysteresis is due to the energy barrier related to the nucleation and growth process, by improving the geometric and magnetic compatibility between the phases the thermal hysteresis decreases. This is important if the Heuslers are to be considered viable for application. Reducing thermal hysteresis increases the energy savings in cyclical applications as well as minimizes fatigue due to crack formation during cycles of transformation.[8]

**IV. Conclusions**

Thermal, structural and magnetic properties of $Ni_2Mn_{1-x}Cu_xGa_{0.84}Al_{0.16}$ Heusler alloys were studied. As Mn is replaced by Cu, the martensitic transition temperature rises while the magnetic transition temperature decreases. A valley-like behavior in the Cu content dependence of the thermal hysteresis was identified and for this material the minimum in $\Delta H$ was found to occur at the Cu content x = 0.25. The geometric non-linear theory of martensite was applied to the structural data and provided invaluable insight. The valley-like behavior of the hysteresis coincides with a region of the phase diagram where the parameter $\lambda_2$ is closer to 1. In addition, a comprehensive analysis of results previously published in the literature on the full Heusler alloys $Ni_2Mn_{1-x}Cu_xGa$, $Ni_{2+x}Mn_{1-x}Ga$, $Ni_2Mn_{1-x}Cr_xGa$ as well as the half Heusler alloys $NiMn_{1-x}In_x$, $NiMn_{1-x}Al_x$ and $NiMn_{1-x}In_x$ reveals that all compounds also show a minimum of the thermal hysteresis width when Mn is replaced by different ions. Understanding and tuning the thermal hysteresis width is a pressing issue if magnetocaloric and shape memory Heusler alloys are to be employed in future applications. In this context, our work establishes a gateway for compositional engineering and exploitation of Heusler and half Heusler with minimum fatigue and minimum thermal hysteresis.

Acknowledgements

A.A.M. is supported by a graduate grant from the Brazilian agency CAPES (Coordenaçãao de Aperfeiçoamento de Pessoal de Nível Superior). L.G. and A. M. G. were supported by CNPq (Conselho Nacional de Desenvolvimento Científico e Tecnológico), projects 305021/2017-6 and 424688/2018-2. L. G. acknowledges financial support from FAPERJ (Fundação Carlos Chagas Filho de Amparo a Pesquisa do Estado do Rio de Janeiro), Projects E-26/202.820/2018 and E-26/010.101136/2018. H. G. and R. D. J. acknowledge the support of NSF (DMREF-1629026), ONR (N00014-18-1-2766) and a Vannevar Bush Faculty Fellowship. L.F.C is supported by the UK EPSRC project number EP/P511109/1 and Innovate UK FlexMag: project 105541 or 32645.

**Figures:**

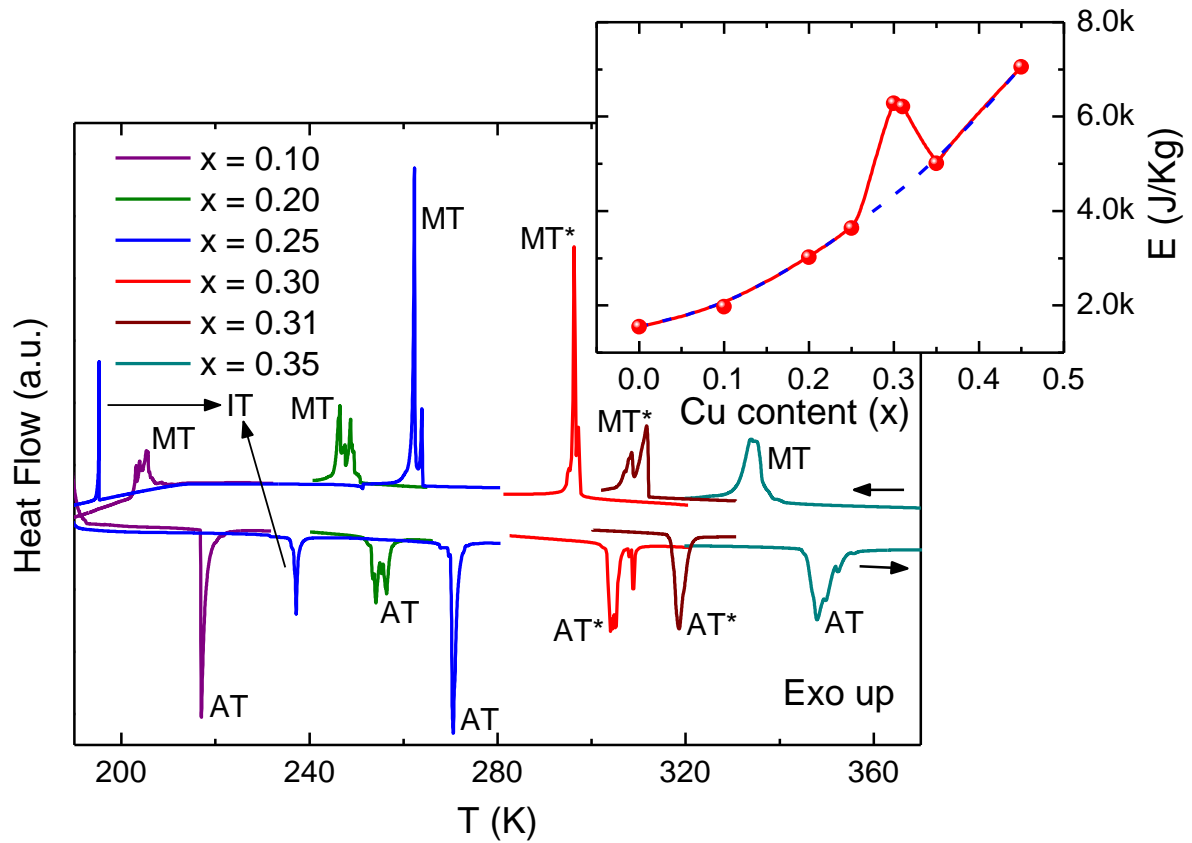

**Figure 1.** Heat flow measurements for $Ni_2Mn_{1-x}Cu_xGa_{0.84}Al_{0.16}$, with x = 0.10, 0.20, 0.25, 0.30, 0.31 and 0.35. Results shown in arbitrary units for better visualization. IT stands for intermartensitic transition, MT and AT are martensitic and austenitic transitions, respectively, where both phases are ferromagnetic, and MT* as well as AT* are martensitic and austenitic transitions, respectively, from paramagnetic to ferromagnetic phase. The heat flow curve for x = 0.45 was not shown because the transition temperature is close to 500 K. Inset: Transition enthalpy for the first order transitions calculated from the heat flow measurement, as a function of Cu content. The dashed line represents the expected enthalpy variation without a magnetostructural transition.

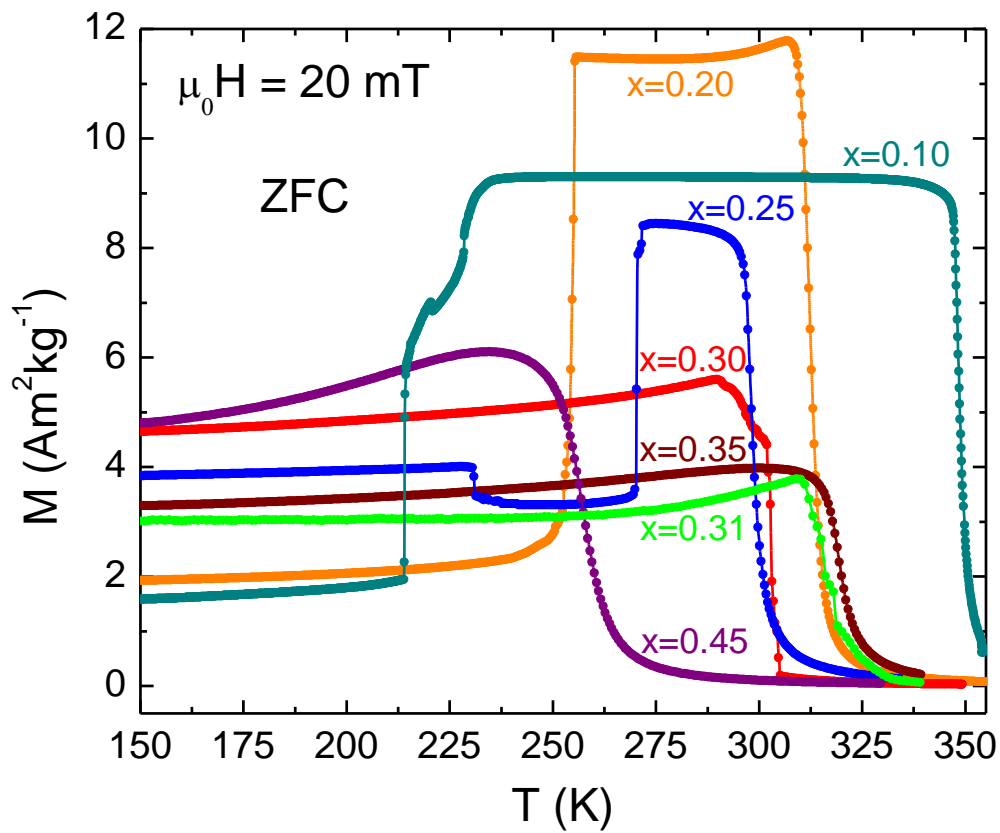

**Figure 2.** Temperature dependence of the magnetization of $Ni_2Mn_{1-x}Cu_xGa_{0.84}Al_{0.16}$, with $x = 0.10$, $0.20$, $0.25$, $0.30$, $0.35$ and $0.45$. The compounds with $x = 0.10$, $0.20$ and $x = 0.25$ display a magnetic transition in the austenite phase, while for $x = 0.35$ and $x = 0.45$ the magnetic transition occurs in the martensite phase. In the case of $x = 0.3$ and $0.31$ there is a MST from the martensite to the austenite phase.

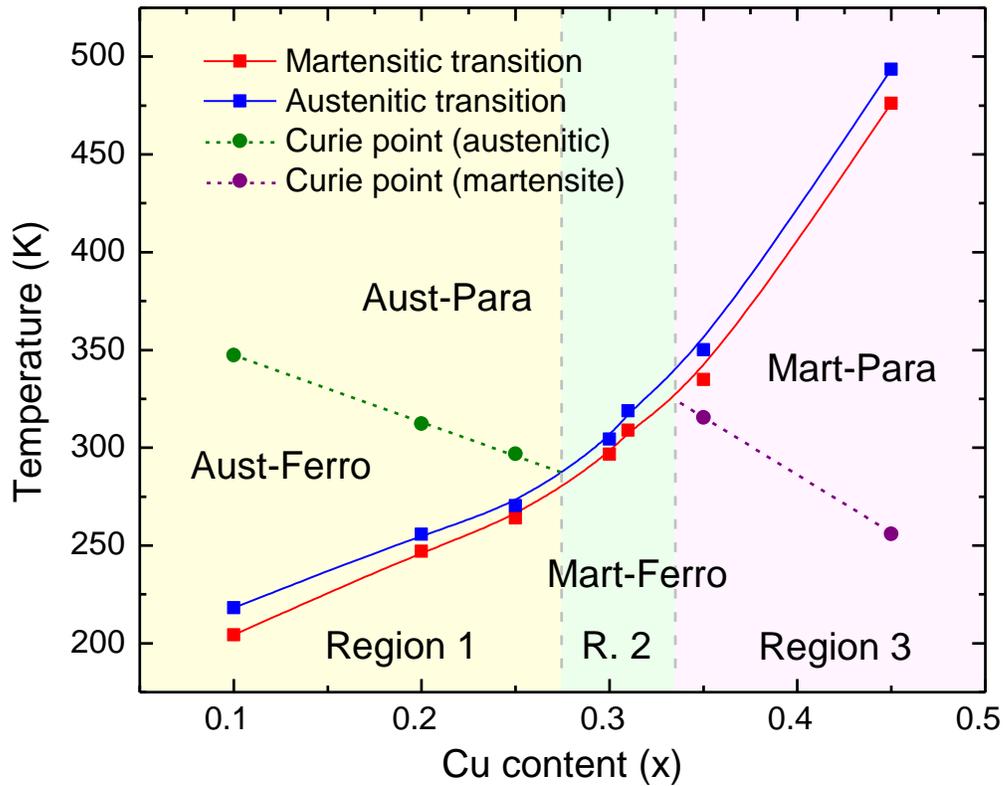

**Figure 3.** Phase diagram of the Ni$_2$Mn$_{1-x}$Cu$_x$Ga$_{0.84}$Al$_{0.16}$ alloys plotted as function of the Cu content, x. The regions with austenite and martensite phase as well as ferromagnetism and paramagnetism are represented. The regions 1, 2 and 3 represent, respectively, the occurrence of magnetic transition in the austenite phase, the MST and the magnetic transition in the martensite phase.

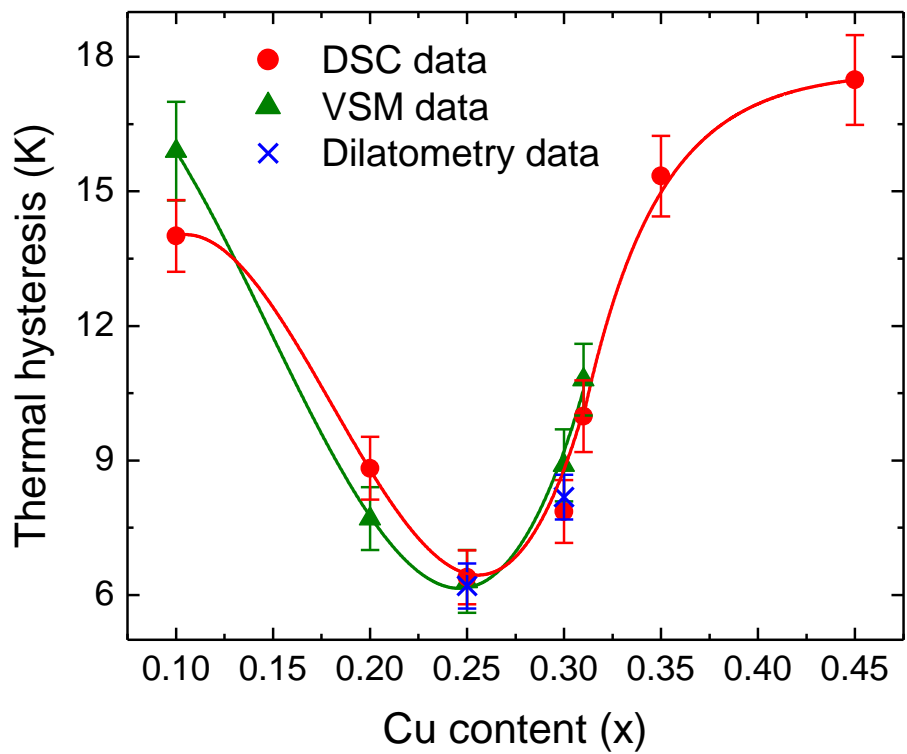

**Figure 4.** Thermal hysteresis width, ΔH, plotted as a function of the Cu content in Ni$_2$Mn$_{1-x}$Cu$_x$Ga$_{0.84}$Al$_{0.16}$ alloys. Results obtained from DSC, magnetization and dilatometry.

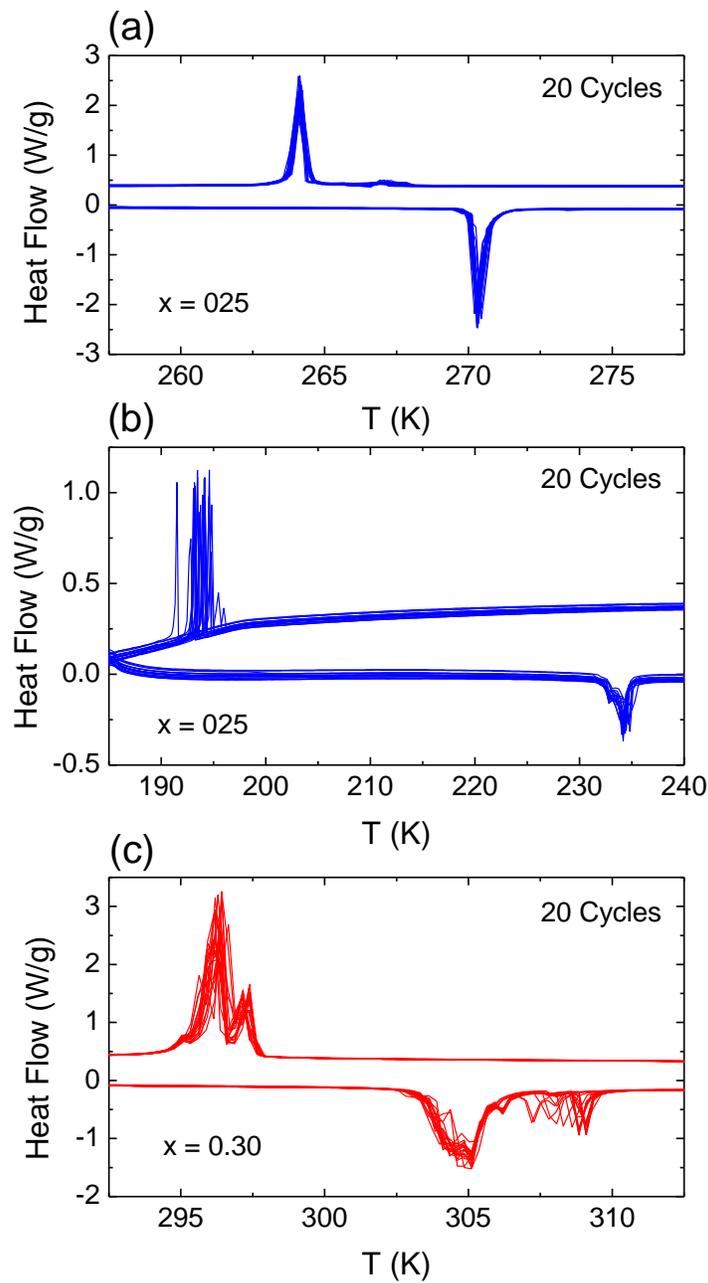

**Figure 5.** Thermal cycles of heat flow around the (a) martensitc and (b) intermartensitic transformation for x = 0.25 and around the (c) MST for 0.30. In each case, 20 cycles where made in order to study the transformation reproducibility and stability.

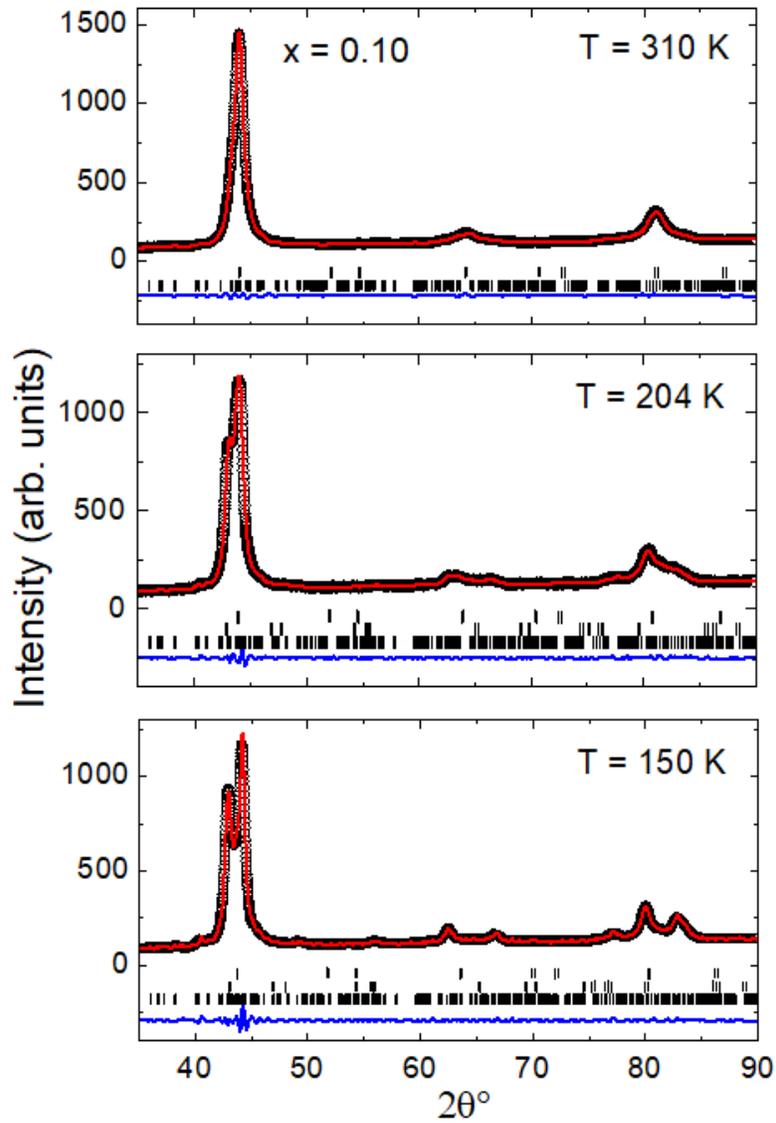

**Figure 6.** X-ray powder diffraction results, refined curve, difference between experimental and calculated data, and Bragg positions for the Ni$_2$Mn$_{1-x}$Cu$_x$Ga$_{0.84}$Al$_{0.16}$, with x = 0.10. The measurements carried out at 310 K, corresponding to a pure austenite phase, at 150 K, corresponding to a pure martensite phase, and at 204 K, where there is a coexistence of both phases.

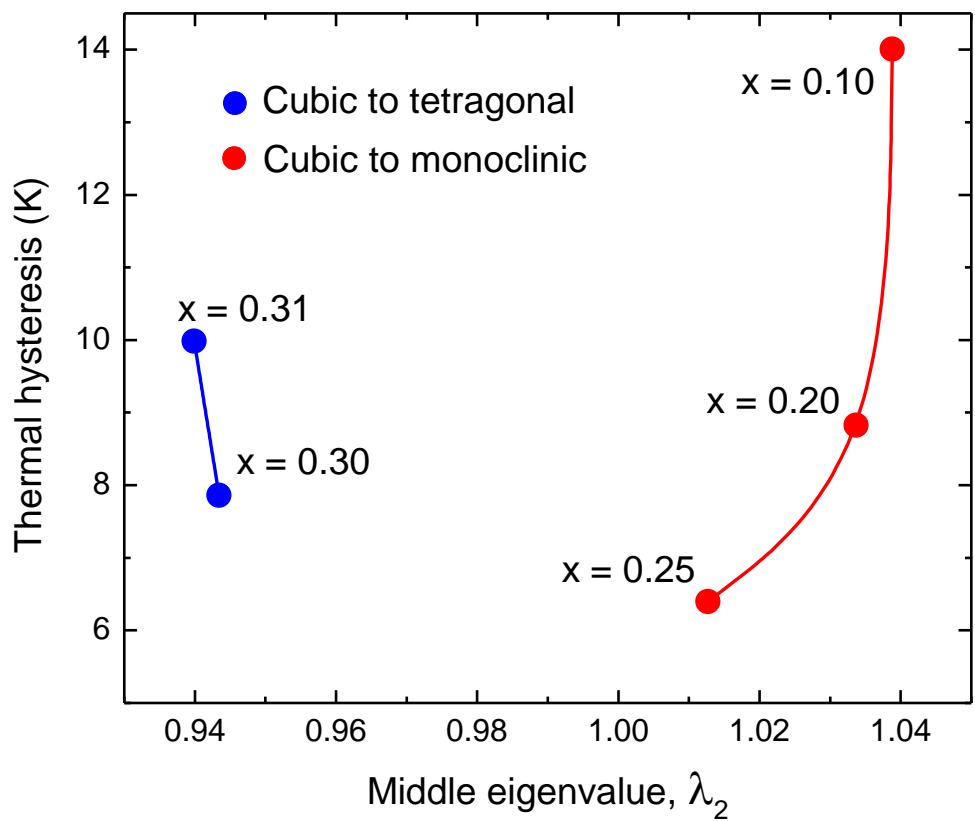

**Figure 7.** Thermal hysteresis, obtained from DSC measurements, as a function of the middle eigenvalue, $\lambda_2$.

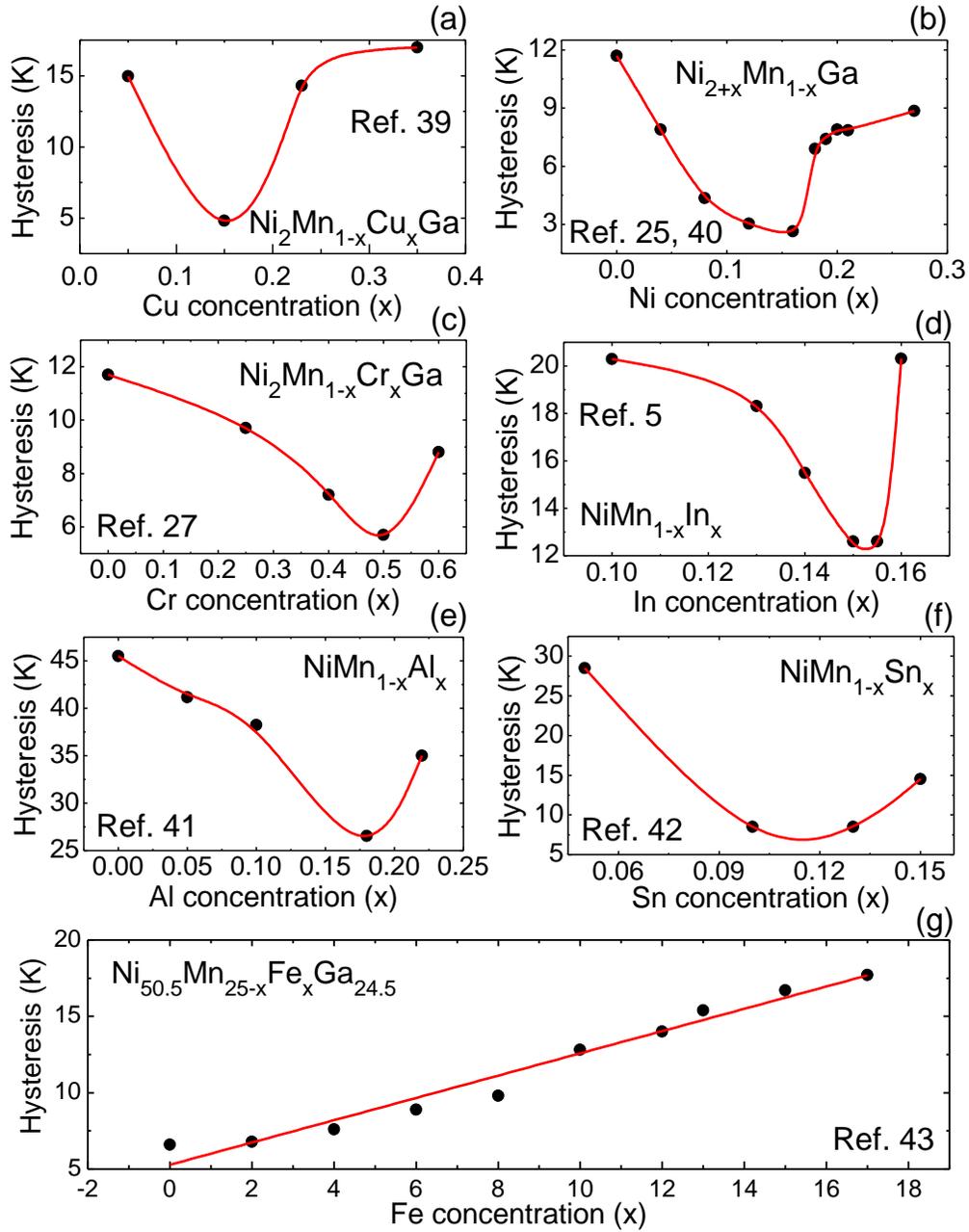

**Figure 8.** Thermal hysteresis obtained from the literature for different materials.

**Table I:** EDS results with the actual composition of the different elements in $Ni_2Mn_{1-x}Cu_xGa_{0.84}Al_{0.16}$ alloys.

| Nominal Cu content, x | Nominal composition of the alloy | EDS Results | | | | |
|---|---|---|---|---|---|---|
| | | Ni (%) | Mn (%) | Cu (%) | Ga (%) | Al (%) |
| 0.10 | $Ni_{49.7}Mn_{20.9}Cu_{2.7}Ga_{24.8}Al_{1.8}$ | 49.4 | 23.9 | 2.7 | 23.9 | 1.9 |
| 0.20 | $Ni_{49.5}Mn_{18.5}Cu_{5.4}Ga_{24.7}Al_{1.8}$ | 48.5 | 19.5 | 5.2 | 24.4 | 2.5 |
| 0.25 | $Ni_{49.5}Mn_{17.4}Cu_{6.7}Ga_{24.7}Al_{1.8}$ | 48.2 | 18.5 | 6.6 | 23.9 | 2.8 |
| 0.30 | $Ni_{49.4}Mn_{16.2}Cu_{8.0}Ga_{24.6}Al_{1.8}$ | 49.4 | 17.3 | 8.1 | 22.8 | 2.3 |
| 0.31 | $Ni_{49.3}Mn_{15.9}Cu_{8.3}Ga_{24.6}Al_{1.8}$ | 48.4 | 16.9 | 8.0 | 24.0 | 2.7 |
| 0.35 | $Ni_{49.3}Mn_{15.0}Cu_{9.3}Ga_{24.6}Al_{1.8}$ | 50.7 | 16.4 | 9.6 | 21.6 | 1.7 |
| 0.45 | $Ni_{49.1}Mn_{12.6}Cu_{12.0}Ga_{24.5}Al_{1.8}$ | 46.9 | 13.5 | 12.0 | 24.4 | 3.2 |

**Table II:** Lattice parameters and middle eigenvalues $\lambda_2$ of $Ni_2Mn_{1-x}Cu_xGa_{0.84}Al_{0.16}$ alloys in the cubic to monoclinic and cubic to tetragonal transformations.

| Cu (x) | Phase transformation | Martensite: | | | | Austenite: | $\lambda_2$ |
|---|---|---|---|---|---|---|---|
| | | $a$ (Å) | $b$ (Å) | $c$ (Å) | $\beta$ (degree) | $a$ (Å) | |
| 0.10 | Cubic to Monoclinic | 4.1872 | 5.5805 | 20.8750 | 90.160 | 5.6809 | 1.0388 |
| 0.20 | Cubic to Monoclinic | 4.2930 | 5.7097 | 20.7054 | 89.027 | 5.6656 | 1.0337 |
| 0.25 | Cubic to Monoclinic | 4.2417 | 5.6791 | 20.7314 | 89.402 | 5.7900 | 1.0127 |
| 0.30 | Cubic to Tetragonal | 3.8735 | 3.8735 | 6.4564 | 90 | 5.8066 | 0.9434 |
| 0.31 | Cubic to Tetragonal | 3.8517 | 3.8517 | 6.4545 | 90 | 5.7954 | 0.9399 |